\documentclass[12pt,a4,portrait]{article}
\usepackage{latexsym}
\usepackage{hyperref}
\usepackage{graphicx}

\newcommand{\be}{\begin{equation}}
\newcommand{\ee}{\end{equation}}
\newcommand{\beq}{\begin{eqnarray}}
\newcommand{\eeq}{\end{eqnarray}}
\newcommand{\bed}{\begin{displaymath}}
\newcommand{\eed}{\end{displaymath}}
\newcommand{\bc}{\begin{center}}
\newcommand{\ec}{\end{center}}
\newcommand{\bi}{\begin{itemize}}
\newcommand{\ei}{\end{itemize}}
\newcommand{\bn}{\begin{enumerate}}
\newcommand{\en}{\end{enumerate}}

\newcommand{\rw}{ {\rm w} }

\begin{document}

\title{Remarks on (super-)accelerating cosmological models \\ in general scalar-tensor gravity}
\author{Laur J\"arv\thanks{laur.jarv@ut.ee}, Piret Kuusk\thanks{piret@fi.tartu.ee} \\
 Institute of Physics, University of Tartu, Riia 142, Tartu 51014, Estonia \\ \\
 Margus Saal\thanks{margus@fi.tartu.ee} \\
 Tartu Observatory, T\~oravere 61602, Estonia}
\date{}
\maketitle

\begin{abstract}
We consider Friedmann-Lema\^{\i}tre-Robertson-Walker
cosmological models in the framework of general scalar-tensor theories of gravity (STG) with arbitrary coupling functions, set in the Jordan frame.
First we describe the general properties of the phase space 
in the case of barotropic matter fluid and scalar field potential for 
any spatial curvature (flat, spherical, hyperbolic). 
Then we address the question under which conditions
 epochs of accelerated and super-accelerated expansion 
are possible in STG. For flat models filled with dust matter (and vanishing potential) 
we give a necessary condition on the coupling function of the scalar field 
which must be satisfied to allow acceleration and super-acceleration. This is illustrated by a specific example.
\end{abstract}


\section{Introduction}

The last decade has produced an abundance of cosmological precision data, 
leading to surprising results and implications.
Approching the statistics of observations by more relaxed priors suggests that 
the expansion of the Universe as measured by the scale factor $a$ is not only
accelerating ($\frac{\ddot{a}}{a}>0$), but might be also about to enter 
into a super-accelerating phase ($\dot{H}=\frac{\ddot{a}}{a} - \frac{\dot{a}^2}{a^2}>0$),
sometimes dubbed as ``crossing the phantom divide'' \cite{phantom_obs}.
The latter possibility can not be accommodated 
in the cosmological Concordance Model based on the Einstein equations with a cosmological constant in the framework of general relativity (GR).  
If one prefers to play within the traditional GR, then the onset of super-acceleration 
can be invoked by adding another matter component 
with unusual ``phantom'' properties \cite{phantom_GR}. 
An alternative explanation would require superseding GR by a more general theory of gravitation, examples of super-accelerating 
 solutions have been studied, for instance, in the context of
f(R) \cite{acc_fR} and scalar-tensor \cite{acc_stg} theories.

Scalar-tensor theories of gravitation (STG) employ a scalar field $\Psi$ besides the usual spacetime metric tensor $g_{\mu \nu}$ to describe the gravitational interaction. Scalar field is in the role of a variable gravitational ``constant'', leaving tensorial metric field and its geodesics to act as trajectories of freely falling particles as in GR. 
In general, STG form a collection of theories which contain two functional degrees of freedom, a coupling function $\omega(\Psi)$ and a scalar potential $V(\Psi)$. Each distinct functional form of these two functions gives us a distinct theory of gravitation together with its field equations. It is of considerable interest to determine which 
members of this family of theories allow solutions (model Universes) exhibiting 
periods of accelerating and super-accelerating expansion without introducing any 
unusual matter component.

The study of global properties of solutions can be greatly facilitated by 
the mathematical methods of dynamical systems and phase space.
Several previous detailed studies which have considered STG cosmology as a dynamical system have focused upon examples with specific coupling functions and potentials \cite{stg_dynsys}. 
The main properties of the corresponding general phase space geometry were outlined by Faraoni \cite{faraoni1} and us \cite{meie4}. 

The plan of the paper is the following. Section 2 introduces STG field equations for homogeneous and isotropic cosmological models. In section 3 we describe the phase space in the most general case: one barotropic matter fluid component,
non-vanishing scalar field potential 
and arbitrary spatial geometry (flat, spherical, hyperbolic), thus generalizing the results of previous studies \cite{faraoni1, meie4}.
In section 4 we investigate the conditions under which accelerated and super-accelerated expansion is possible, and also when do the solutions enter
or leave the epoch of accelerated and super-accelerated expansion.
In the simplest and phenomenologically most relevant case of dust matter, 
vanishing potential, and flat 
spatial geometry ($k=0$) we give a necessary condition on the coupling function $\omega(\Psi)$ which must be satisfied for
 acceleration and super-acceleration to be possible at all.
These considerations are illustrated by an example of a particular STG where
some solutions undergo a phase of super-acceleration while some solutions do not.

\section{The equations of scalar-tensor cosmology} 

We consider  a general scalar-tensor theory
in the Jordan frame given by the action functional
    \beq \label{jf4da}
S  = \frac{1}{2 \kappa^2} \int d^4 x \sqrt{-g}
        	        \left[ \Psi R(g) - \frac{\omega (\Psi ) }{\Psi}
        		\nabla^{\rho}\Psi \nabla_{\rho}\Psi
                  - 2 \kappa^2 V(\Psi)  \right]
                   + S_{m}(g_{\mu\nu}, \chi_m) \,.
\eeq 
Here $\omega(\Psi)$ is a coupling function and $V(\Psi)$ is a
scalar potential, $\nabla_{\mu}$ 
denotes the covariant derivative with respect to the metric 
$g_{\mu\nu}$, $\kappa^2$ is the non-variable part of
the gravitational constant, and 
$S_{m}$ is the matter contribution to the action 
as all other fields are included in $\chi_m$.
In order to keep the effective gravitational constant $\frac{\kappa^2}{\Psi}$ positive
we assume that $0 < \Psi < \infty$.

The field equations for the Friedmann-Lema\^{\i}tre-Robertson-Walker
(FLRW) line element 
\be
ds^2=-dt^2 + a(t)^2 \left( \frac{dr^2}{1-kr^2} + r^2 (d\theta ^2 + \sin ^2 \theta d\varphi ^2)\right)
\ee
with curvature parameter $k= 0$ (flat), $+1$ (spherical), $-1$ (hyperbolic), and perfect barotropic fluid 
matter, $p=\rw \rho$, read 
\beq 
\label{00}
H^2 &=& 
- H \frac{\dot \Psi}{\Psi} 
+ \frac{1}{6} \frac{\dot \Psi^2}{\Psi^2} \ \omega(\Psi) 
+ \frac{\kappa^2}{3} \frac{ \rho}{\Psi} 
+ \frac{\kappa^2}{3} \frac{V(\Psi)}{\Psi} - K \,, 
\\ \nonumber \\
\label{mn}
2 \dot{ H} + 3 H^2 &=& 
- 2 H \frac{\dot{\Psi}}{\Psi} 
- \frac{1}{2} \frac{\dot{\Psi}^2}{\Psi^2} \ \omega(\Psi) 
- \frac{\ddot{\Psi}}{\Psi} 
- \frac{\kappa^2}{\Psi} \rw \rho 
+ \frac{\kappa^2}{\Psi} \ V(\Psi)-K \,, 
\\ \nonumber \\
\label{deq}
\ddot \Psi &= & 
- 3H \dot \Psi 
- \frac{1}{2\omega(\Psi) + 3} \ \frac{d \omega(\Psi)}{d \Psi} \  \dot {\Psi}^2 
+ \frac{\kappa^2}{2 \omega(\Psi) + 3} \ (1-3 \rw) \ \rho  \nonumber \\ 
& &\qquad + \frac{2 \kappa^2}{2 \omega(\Psi) + 3} \ \left[ 2V(\Psi) - \Psi \ 
\frac{d V(\Psi)}{d \Psi}\right] \, ,
\eeq 
where $H \equiv \dot{a} / a$, $K=\frac{k}{a^2}$. 
The matter conservation law is the usual 
\beq \label{matter_conservation}
\dot{\rho} + 3 H \ (\rw+1) \ \rho = 0 \,
\eeq
and it is reasonable to assume positive matter energy density, $\rho \geq 0$.

\section{Phase space}

The system (\ref{00})-(\ref{matter_conservation}) is characterized by five variables $\{ \Psi, \dot{\Psi}, H, a, \rho \}$, but
one of them is algebraically related to the others via the Friedmann equation (\ref{00}).
Since the scale factor $a$ is not physically observable, it is reasonable 
to eliminate $K$ by Eq. (\ref{00}). This leads to 
 a phase space spanned by four variables $\{ \Psi, \dot{\Psi}, H, \rho \}$.
By 
defining $\Psi \equiv x, \dot{\Psi} \equiv y$ the dynamical system corresponding to
equations 
(\ref{00})-(\ref{matter_conservation}) can be written
as follows:
\beq
\label{general_dynsys_x}
\dot{x} &=& y \,, \\ 
\label{general_dynsys_y}
\dot{y} &=& -\frac{1}{2\omega(x)+3} \left[ 
\frac{d\omega(x)}{dx} y^2 -\kappa^2 \left( 1-3\rw \right)\rho
+2 \kappa^2 \left( \frac{dV(x)}{dx}x-2V(x) \right)
 \right] -3Hy \,, \\
\label{general_dynsys_H}
\dot{H} &=& \frac{1}{2x(2\omega(x)+3)} \left[ 
\frac{d\omega(x)}{dx} y^2 -\kappa^2 \left( 1-3\rw \right)\rho
+2 \kappa^2 \left( \frac{dV(x)}{dx}x -2V(x) \right)
 \right] \nonumber \\ 
&& \quad - H^2 + H\frac{y}{x} - \omega (x) \frac{y^2}{3x^2} - 
\frac{\kappa^2}{6x} (1+ 3 \rw) \rho + \frac{\kappa^2 V(x)}{3x} \,,  \\ 
\dot{\rho} &=& -3 H (1+\rw) \rho \,.
\label{general_dynsys_rho}
\eeq

The phase space may be imagined as a four dimensional box filled by 
the spaghetti of one dimensional 
trajectories (orbits of solutions) which do not intersect with each other except for special points known
as fixed (critical, equilibrium) points. 
As the curvature invariants of FLRW metric are proportional to $H$ and $\dot{H}$ 
the phase space boundaries $|H| \rightarrow \infty$, $|\rho| \rightarrow \infty$,
and $|\dot{\Psi}| \rightarrow \infty$ generically entail a spacetime singularity.
Analogously, the limit $\Psi \rightarrow 0$ in general implies 
diverging $|H|$ or $|\dot{H}|$ and poses a spacetime singularity, obstructing
the solutions from safely passing from positive to negative values of 
$\Psi$ (from ``attractive'' to ``repulsive'' gravity).
The limit $\Psi \rightarrow \infty$ does not call forth a spacetime singularity,
however, the gravitational ``constant'' vanishes.

Within this box there could also be singular hypersurfaces perpendicular 
to the $\Psi$ axis,
depending on the form of $\omega$ and $V$.
So, in general terms the limit $V \rightarrow \infty$ renders the system singular, 
and also $2\omega + 3 \rightarrow 0$ implies 
$|\dot{H}| \rightarrow \infty$ with the same conclusion that 
passing through $\omega(\Psi)=-\frac{3}{2}$ 
(corresponding to the change of the sign of the scalar field kinetic
term in the Einstein frame action)
would entail a space-time singularity and is impossible. 
The limit $\frac{1}{2\omega + 3} \rightarrow 0$ is also marred by a singularity,
unless simultaneously
\be \label{GR_limit}
\dot{\Psi} \rightarrow 0, \qquad \omega \dot{\Psi}^2 \rightarrow 0, \qquad
\frac{1}{(2\omega + 3)^2} \frac{d\omega}{d\Psi} \rightarrow {\rm finite} \, .
\ee
The latter situation is particularly interesting,
since in this limit the system coincides with the FLRW equations of
general relativity \cite{meie4}.

The trajectories corresponding to the flat FLRW geometry ($k=0$) lie on 
the 3-surface
\be
\label{Friedmann_surface}
\mathcal{F}: F(x,y,H,\rho)\equiv H^2 + H\frac{y}{x} - \frac{y^2}{6 x^2} \ \omega(x) 
- \frac{\kappa^2 \rho}{3x} -\frac{\kappa^2 V(x)}{3x} =0\,.
\ee
due to the constraint (\ref{00}). The trajectories corresponding to spherical and hyperbolic models remain on either 
side of this surface.
In principle the geometry of the 3-surface $\mathcal{F}$ in the 4-dimensional
phase space
is rather complicated to visualize,
but a few general characteristics can still be given.
We may write Eq. (\ref{Friedmann_surface}) as
\be
\frac{\left(H + \frac{y}{2x}\right)^2}{\frac{\kappa^2(\rho+V)}{3x}}
- \frac{y^2}{\frac{4 \kappa^2 x (\rho+V)}{2\omega+3}} = 1 \, ,
\ee
which for fixed $\rho$ and $x$ can be recognized as describing 
familiar conic sections:
1) for $\rho+V>0$, $2\omega+3>0$ a hyperbola 
on the $(H + \frac{y}{2x},y)$ plane, 
2) for $\rho+V>0$, $2\omega+3<0$ an ellipse 
also on the $(H + \frac{y}{2x},y)$ plane,
while
3) for $\rho+V<0$, $2\omega+3>0$ a hyperbola 
on the $(y, H + \frac{y}{2x})$ plane. 
The case 
4) $\rho+V<0$, $2\omega+3<0$ is not realized as 
real solutions are absent.
This result establishes that the intersection of the 3-surface $\mathcal{F}$
with the (fixed $\rho$, fixed $x$) 2-plane is constituted in either
one piece (ellipse) or two pieces (hyperbola). 
Thus in case 1) the allowed phase space is divided into two separate regions,
the ``upper'' region where $H+\frac{y}{2x}>0$ and 
the ``lower'' region
 where $H+\frac{y}{2x}<0$, and
there is no way the trajectories can travel from one region to another. 
In case 2) these two regions meet along a 2-surface where 
$H+\frac{y}{2x}=0$, and the trajectories can in principle cross
from one region to another. 
In case 3) there are again two separate parts, now characterized by
$y>0$ and $y<0$, respectively.
At first it may be difficult find a direct physical interpretation 
for the quantity $H + \frac{y}{2x}$ that characterizes the 
``upper'' and ``lower'' region in cases 1) and 2), 
but it turns out that this combination is equal to the Hubble parameter in
the Einstein frame \cite{faraoni3, meie4}, and thus the ``upper''
region corresponds to universes which expand 
in the Einstein frame, 
while 
the ``lower'' region has universes 
which contract in the Einstein frame.

Related information can be also established by another approach.
With general $k$ we may solve the Friedmann constraint, Eq. (\ref{00}), for $H$
and then the condition for all variables to be real valued 
imposes an inequality
\be
\label{y_allowed_inequality}
 \left( 2\omega(x)+3 \right) \frac{y^2}{12x^2} 
+ \frac{\kappa^2 (\rho + V(x))}{3x} \ge K \, .
\ee
In terms of physics 
this inequality can be interpreted as a restriction on the allowed 
values of $y$. Table \ref{allowed_regions} summarizes the situation. 
By the allowed range for a given value of $k$ 
we mean that if $y$ satisfies the inequality listed, 
then it is possible to find a real-valued $a$ which fits the Friedmann equation.
Thus for $k=0$ and $k=+1$ there is no restriction in case 1), while
the case 
4) is completely ruled out
since no
real solutions compatible with the Friedmann constraint exist. For
$k=-1$ there are no restrictions.

Similarly, solving the Friedmann constraint for $y$ leads to 
another inequality,
\be
\label{H_allowed_inequality}
(2\omega(x)+3)H^2 - \frac{2 \kappa^2 \omega(x)}{3x} (\rho+V(x)) \ge -2 K \omega(x) \, ,
\ee
which can be interpreted as a restriction on the allowed values of $H$
(given also in Table \ref{allowed_regions}). 
Analogously, once $\omega(x)$ and $V(x)$ are specified, 
we may get a third inequality from solving the Friedmann constraint for $x$
as well.
\begin{table}[t] 
\begin{center}
\begin{tabular}{llcccclcl}
&  & \qquad &  & \quad & Allowed range of $\dot \Psi$ && Allowed range of $H$ \\
& & & & & for $k=0, +1$ && for $k=0,-1$
\vspace{2mm}\\
\hline \\
1a) & $\rho+V \geq 0$ && $0 \leq \omega \leq \infty$ 
&& $0 \leq {\dot\Psi}^2 \leq \infty$ 
&& $\frac{2\kappa^2\omega(\rho+V)}{3\Psi(2\omega+3)} \leq H^2 \leq \infty$
\vspace{2mm}\\
1b) & $\rho+V \geq 0$ && $-\frac{3}{2} \leq \omega \leq 0$ 
&& $0 \leq {\dot\Psi}^2 \leq \infty$ 
&& $0 \leq H^2 \leq \infty$
\vspace{2mm}\\
2) & $\rho+V > 0$ && $-\infty \leq \omega \leq -\frac{3}{2}$ 
&& $0 \leq {\dot\Psi}^2 \leq \frac{4 \kappa^2 (\rho+V) \Psi}{|2 \omega+3|}$
&& $0 \leq H^2 \leq \frac{2\kappa^2\omega(\rho+V)}{3\Psi(2\omega+3)}$
\vspace{6mm}\\
&  & \qquad &  & \quad & Allowed range of $\dot \Psi$ && Allowed range of $H$ \\
& & & & & for $k=0, +1$ && for $k=0,+1$
\vspace{2mm}\\
\hline \\
3a) & $\rho+V \leq 0$ && $0 \leq \omega \leq \infty$ 
&& $\frac{4 \kappa^2 |\rho+V| \Psi}{2 \omega+3} \leq {\dot\Psi}^2 \leq \infty$
&& $0 \leq H^2 \leq \infty$
\vspace{2mm}\\
3b) & $\rho+V \leq 0$ && $-\frac{3}{2} \leq \omega \leq 0$ 
&& $\frac{4 \kappa^2 |\rho+V| \Psi}{2 \omega+3} \leq {\dot\Psi}^2 \leq \infty$ 
&& $\frac{2\kappa^2\omega(\rho+V)}{3\Psi(2\omega+3)} \leq H^2 \leq \infty$
\vspace{2mm}\\
4) & $\rho+V < 0$ && $-\infty < \omega < -\frac{3}{2}$ 
&& --
&& --
\vspace{4mm}\\
\hline
\end{tabular}
\end{center}
\caption{For certain $k$ 
the Friedmann equation constrains the values of 
$\dot{\Psi}\equiv y$ (\ref{y_allowed_inequality}) and 
$H$ (\ref{H_allowed_inequality}).}
\label{allowed_regions}
\end{table}

\section{Acceleration and super-acceleration}

In the four dimensional phase space there could be regions where
the trajectories exhibit super-accelerating behavior, marked by the 
condition
\beq \label{sacc}
S(x,y,H,\rho) & \equiv & \frac{1}{2x(2\omega(x)+3)} \left[ 
\frac{d\omega(x)}{dx} y^2 -\kappa^2 \left( 1-3\rw \right)\rho
+2 \kappa^2 \left( \frac{dV(x)}{dx}x -2V(x) \right)
 \right] \nonumber \\ 
&& \qquad \quad - H^2 + H\frac{y}{x} - \omega (x) \frac{y^2}{3x^2} - 
\frac{\kappa^2}{6x} (1+ 3 \rw) \rho + \frac{\kappa^2 V(x)}{3x} >0 \, ,
\eeq
and surrounded by regions of accelerated expansion, delineated by
\beq \label{acc}
A(x,y,H,\rho) & \equiv & \frac{1}{2x(2\omega(x)+3)} \left[ 
\frac{d\omega(x)}{dx} y^2 -\kappa^2 \left( 1-3\rw \right)\rho
+2 \kappa^2 \left( \frac{dV(x)}{dx}x -2V(x) \right)
 \right] \nonumber \\ 
&& \qquad \qquad \qquad + H\frac{y}{x} - \omega (x) \frac{y^2}{3x^2} - 
\frac{\kappa^2}{6x} (1+ 3 \rw) \rho + \frac{\kappa^2 V(x)}{3x} >0 \,. 
\eeq
For a cursory comparison with general relativity let us recall that 
in GR super-acceleration requires matter with barotropic index $\rw<-1$, while acceleration demands $\rw<-\frac{1}{3}$. 
Cosmological constant behaves as a barotropic fluid with $\rw=-1$.
Eqs. (\ref{sacc}), (\ref{acc}) readily reveal that 
identical conditions are recovered in STG at the GR limit (\ref{GR_limit}), where the fixed value of the potential is 
read as the cosmological constant, and the Friedmann constraint
(\ref{00}) should be taken into account along with (\ref{sacc}).

However, away from the GR limit, new possibilities occur.
First, irrespective of the matter content, the scalar field
itself may trigger accelerated and super-accelerated expansion, in the 
domain where $\omega<0$, or $\frac{1}{2x(2\omega+3)} \frac{d\omega}{d\Psi}>0$.
Second, acceleration and super-acceleration may also occur
for matter with $\rw > \frac{1}{3}$ in the domain where $2\omega+3 > 0$,
or for  $\rw < \frac{1}{3}$ in the domain where $2\omega+3 < 0$.
Third, the overall effect of the potential is considerably more complicated
than that of a simple cosmological constant,  
depending on the derivative $\frac{dV}{d\Psi}$ as well as the sign of $2\omega+3$, and also
 implying a possibility that a constant negative potential may
lead to super-acceleration, provided that $2\omega+3 > 0$.

The region of super-acceleration is bounded by the 3-surface 
$\mathcal{S}: S(x,y,H,\rho)=0$. The circumstance whether the 
trajectories enter this region can be read off from the 
scalar product of the gradient normal to $\mathcal{S}$ and 
the tangent vector of 
the phase flow $T^i = (\dot{x}, \dot{y}, \dot{H}, \dot{\rho})$, 
 namely $\nabla_i S \cdot T^i \Big|_{\mathcal{S}} >0$. 
A completely analogous condition arises for the surface $\mathcal{A}=0$
bounding the region of acceleration.

These results, expressed in full generality,
hint ample possibilities for acceleration and super-acceleration,
but in order to come up with more exact conditions
one has to narrow down the scope a bit.
Therefore let us focus upon the physically most interesting
case of spatially flat ($k=0$) universe filled with dust matter 
($\rw=0$) and vanishing potential. In this case Eq. (\ref{mn}) 
can be written as
\beq
\dot{H} &=& - \frac{H^2}{2} -\frac{5}{12}\frac{\omega \dot{\Psi}^2}{\Psi^2} - 
\frac{\ddot\Psi}{2\Psi} - \frac{\kappa^2 \rho}{3 \Psi} \nonumber \\
&=& -2H^2 + \frac{\kappa^2 \omega \rho}{3 \Psi (2\omega+3)} 
+ \frac{\dot{\Psi}^2}{2\Psi^2} \left(\frac{\Psi}{2\omega+3}\frac{d\omega}{d\Psi} - \frac{\omega}{3} \right) \, .
\label{sacc_dust}
\eeq
Here the first line informs that super-acceleration 
is only possible, if $\omega<0$, or $\ddot{\Psi}<0$.
From the second line which has taken Eq. (\ref{deq}) into account,
we can read off a neccessary condition on the form of the 
coupling function $\omega$ for super-acceleration to be possible
\be \label{c_condition}
\mathcal{C} = \frac{\Psi}{2\omega+3}\frac{d\omega}{d\Psi} - \frac{\omega}{3} > 0 \, ,
\ee
assuming $2\omega+3>0$. The reason is that if $\omega>0$ the Fridmann constraint imposes 
$\frac{\kappa^2\omega(\rho)}{3\Psi(2\omega+3)} \leq \frac{H^2}{2}$, 
and the only positive contribution towards super-acceleration
can arise from the third term in 
(\ref{sacc_dust}). The same is true for $-\frac{3}{2} < \omega < 0$ 
since then the second term is negative itself. 
(It is easier to achieve super-acceleration if $2\omega+3<0$,
but this option is not so 
lucrative since in the Einstein frame, where the tensor and scalar  degrees are not mixed, the kinetic energy of the latter is 
negative, and thus problematic \cite{EspositoFarese:2000ij}.)

Note that Eq. (\ref{c_condition}) provides only a neccessary, and
not sufficient condition for super-accelerating solutions to be present 
in a model. More exactly, it states that in the domain of $\Psi$, where (\ref{c_condition}) holds, there may be solutions which undergo 
super-accelerated expansion. In the domain of $\Psi$, where (\ref{c_condition}) does not hold, super-acceleration is not possible.
Therefore, given a zoo of all possible forms of $\omega$, 
it
can be used to filter out and discard from further investigation those 
forms of $\omega$, which are decidedly infertile with respect to super-acceleration.

\begin{figure}[tp]
\begin{center}
\includegraphics[width=8cm, angle=-90]{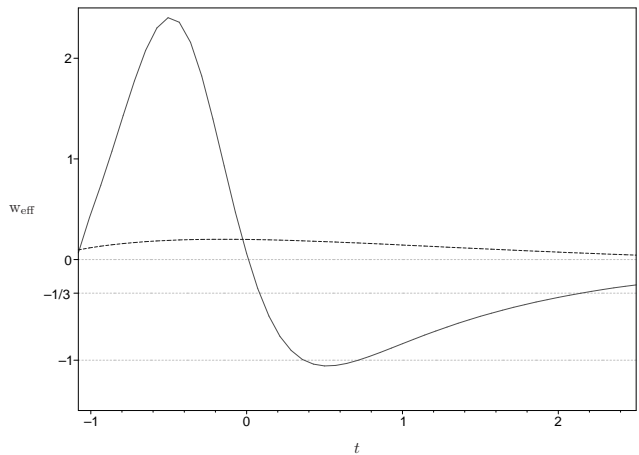}
\end{center}
\caption{Two typical cosmological evolutions in the case of $2\omega+3=\frac{1}{2(1-\Psi)}$, $V(\Psi) \equiv 0$, $\rw=0$, $k=0$: one solution (solid line) goes through
a brief period of super-accelerated expansion (where $\rw_{\rm eff}<-1$), another solution (dashed line) does not.}
\label{example_plots}
\end{figure}

Finally, as an illustration, 
let us consider $2\omega+3=\frac{1}{2(1-\Psi)}$ for example.
Here $2\omega+3>0$ and (\ref{c_condition}) holds in
 the domain $0<\Psi<1$.
 Inspection of
the phase space flow reveals that the trajectories on the 
``upper sheet'' ($H+\frac{\dot{\Psi}}{2\Psi}>0$), where most of the expanding,
$H>0$, models 
lie, belong to two typical classes: either exhibiting a super-accelerating phase or not,
see Fig. \ref{example_plots}.
The dynamics has been characterized by 
the evolution of the effective barotropic index, 
$\rw_{\rm eff} = -1 - \frac{2\dot{H}}{3 H^2}$,
defined as an
analogy to single component barotropic fluid FLRW models in GR. 
In particular, $\rw_{\rm eff}=0$ characterizes the decelerating evolution of
usual dust matter, $\rw_{\rm eff}<-\frac{1}{3}$ is required for acceleration,
while $\rw_{\rm eff}=-1$ corresponds to the ``phantom divide line'' below which 
super-acceleration occurs.

\bigskip
{\bf Acknowledgements}

This work was supported by the Estonian Science Foundation
Grant No. 7185 and by Estonian Ministry for Education and Science
Support Grant No. SF0180013s07. 
MS also acknowledges the 
Estonian Science Foundation Post-doctoral research grant No. JD131.
The authors would like to thank the organizers of 
Baltic-Nordic AGMF Workshop '08, where part of the present paper was presented.

\end{document}